# The UV-SCOPE Mission: Ultraviolet Spectroscopic Characterization Of Planets and their Environments

David R. Ardila[1a], Evgenya Shkolnik[b], John Ziemer[a], Mark Swain[a], James E. Owen[c], Michael Line[b], R. O. Parke Loyd[d], R. Glenn Sellar[a], Travis Barman[e], Courtney Dressing[f], William Frazier[a], April D. Jewell[a], Robert J. Kinsey[a], Carl C. Liebe[a], Joshua D. Lothringer[g], Luz Maria Martinez-Sierra[a], James McGuire[a], Victoria Meadows[h], Ruth Murray-Clay[i], Shouleh Nikzad[a], Sarah Peacock[j], Hilke Schlichting[k], David Sing[l], Kevin Stevenson[m], Yen-Hung Wu[a]

[a]Jet Propulsion Laboratory, California Institute of Technology, [b]Arizona State University, [c]Imperial College London, [d]Eureka Scientific Inc., [e]University of Arizona at Tucson, [f]University of California at Berkeley, [g]Utah Valley University, [h]University of Washington at Seattle, [i]University of California at Santa Cruz, [j]Goddard Space Flight Center, [k]University of California at Los Angeles, [l]The Johns Hopkins University, [m]Applied Physics Laboratory

**ABSTRACT**

UV-SCOPE is a mission concept to determine the causes of atmospheric mass loss in exoplanets, investigate the mechanisms driving aerosol formation in hot Jupiters, and study the influence of the stellar environment on atmospheric evolution and habitability. As part of these investigations, the mission will generate a broad-purpose legacy database of time-domain ultraviolet (UV) spectra for nearly 200 stars and planets.

The observatory consists of a 60 cm, f/10 telescope paired to a long-slit spectrograph, yielding simultaneous, almost continuous coverage between 1203 Å and 4000 Å, with resolutions ranging from 6000 to 240. The efficient instrument provides throughputs > 4% (far-UV; FUV) and > 15% (near-UV; NUV), comparable to *HST/COS* and much better than *HST/STIS*, over the same spectral range. A key design feature is the LiF prism, which serves as a dispersive element and provides high throughput even after accounting for radiation degradation. The use of two delta-doped Electron-Multiplying CCD detectors with UV-optimized, single-layer anti-reflection coatings provides high quantum efficiency and low detector noise. From the Earth-Sun second Lagrangian point, UV-SCOPE will continuously observe planetary transits and stellar variability in the full FUV-to-NUV range, with negligible astrophysical background.

All these features make UV-SCOPE the ideal instrument to study exoplanetary atmospheres and the impact of host stars on their planets. UV-SCOPE was proposed to NASA as a Medium Explorer (MidEx) mission for the 2021 Announcement of Opportunity. If approved, the observatory will be developed over a 5-year period. Its primary science mission takes 34 months to complete. The spacecraft carries enough fuel for 6 years of operations.

**Keywords:** ultraviolet, spectroscopy, exoplanets, transits, astronomy, delta-doping, LiF, time-domain

## 1. INTRODUCTION

Exoplanets outnumber stars in our galaxy and are much more diverse than Solar System planets. Their extreme diversity in mass, radius, stellar irradiation, and composition, presents an immense challenge to our understanding of planet evolution and their potential habitability.

The *Ultraviolet Spectroscopic Characterization Of Planets and their Environments* (UV-SCOPE) mission is a concept that seeks to understand the key physical drivers shaping broad types of exoplanet atmospheres while providing a legacy dataset of simultaneous far-ultraviolet (FUV: 1203 Å - 1560 Å) to near-ultraviolet (NUV, defined for the purposes of UV-SCOPE as 1570 Å to 4000 Å) observations (Figure 1).

---

[1] david.r.ardila@jpl.nasa.gov

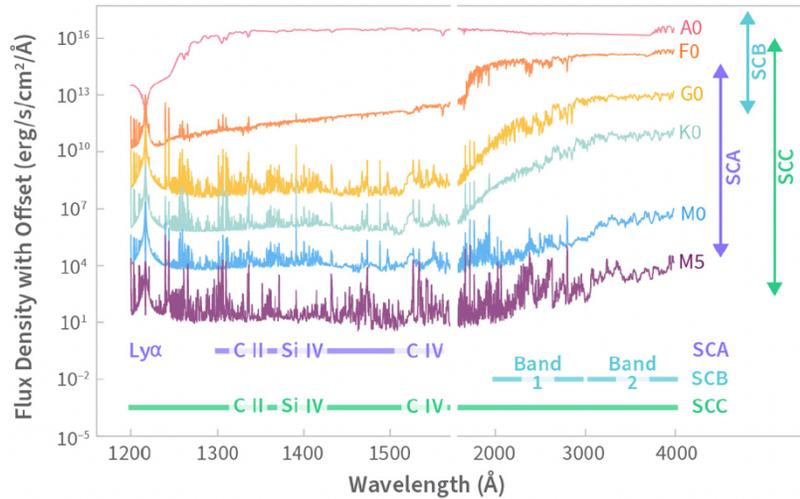

**Figure 1.** UV-SCOPE is designed to provide simultaneous spectroscopic measurements over nearly the complete wavelength range between 1203 Å and 4000 Å. Science topics are organized in Science Cases A, B, and C (SCA, SCB, SCC; Section 2). The figure shows model UV-SCOPE spectra for a range of stellar spectral types. The spectral diagnostics associated with each case are also indicated. UV-SCOPE's observations will generate a rich legacy dataset of stellar spectra and planetary transits, with broad applicability to many other topics of exoplanet science.

The spectral range observed by UV-SCOPE includes emission lines diagnostic of planetary evaporation and lines that can be used to predict the (not directly measurable) stellar flux responsible for that evaporation. It also includes broad molecular lines diagnostic of the formation and composition of clouds in exoplanet atmospheres. In this spectral range, stellar flares and other variability modifies the planet's climate and confounds evidence for life (Figure 2).

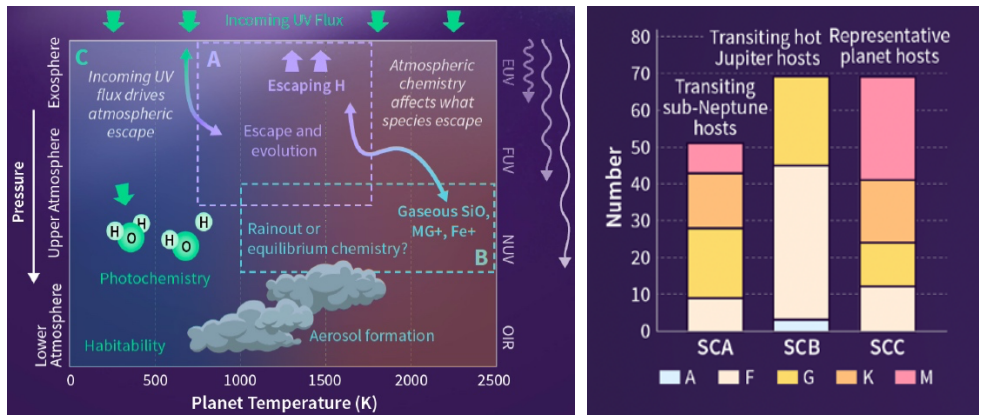

**Figure 2.** UV-SCOPE will study mass loss and chemistry processes over a diverse planet population, as well as the stellar environment of these exoplanets. *Left*: Stellar UV emission powerfully impacts photochemistry and photoevaporation in planetary atmospheres. In turn, planetary atmospheres exhibit strong observable signals in the UV due to high atomic and molecular opacities. *Right*: UV-SCOPE will observe UV transmission spectra of 120

transiting exoplanets orbiting young and old stars with a broad range of spectral types (A through M), plus the spectra of another 69 stars to map the variability, flaring, and evolution of exoplanet high-energy environments.

UV-SCOPE's exoplanet target sample overlaps with the samples that will be observed in the visible and infrared by the James Webb Space Telescope (*JWST*), the Atmospheric Remote-sensing Infrared Exoplanet Large-survey (*ARIEL*), and ground-based extremely large telescopes, providing the missing stellar UV context needed to understand the planets and their potential habitability.

UV-SCOPE will observe from the Earth-Sun second Lagrangian point (L2) where, compared to low-earth orbit (LEO), the UV backgrounds are negligible and it is possible to observe the targets over long, uninterrupted periods (> tens of hours). These continuous observations, performed with broad UV wavelength coverage, are crucial for the analysis and interpretation planetary transits and stellar variability.

## 2. SCIENCE

UV-SCOPE is a direct response to the needs of the US astronomical community as identified in the latest Decadal Survey on Astronomy and Astrophysics (Astro2020) [1]. In the Astro2020 report, the US astronomical community identified the theme *Worlds and Suns in Context*, seeking to build "on revolutionary advances in our observations of exoplanets and stars" and aiming to "understand their formation, evolution, and interconnected nature, and to characterize other solar systems, including potentially habitable analogs to our own." Astro2020 also highlighted specific research needs, such as determining the processes that led to planetary diversity, understanding how habitable environments evolve, and identifying potential signs of life in the context of planetary environments (Astro2020, pg. E-17).

UV-SCOPE's goals match those defined by Astro2020 and are organized around three science cases addressing fundamental questions in exoplanet science (Figure 3):

A. Science Case A (SCA): What causes atmospheric mass loss in planets? What is sculpting the radii of the small-planet population?

   Planetary atmospheres are known to lose mass into space over their lifetimes: For example, one of the leading hypotheses for Venus' water loss and its transformation into an uninhabitable world is that extreme heating by the young Sun's elevated UV radiation drove its atmospheric mass loss [e.g. 2]. In addition to this photoevaporation (PE [3]) driven by the stellar high-energy photons, residual internal heat, leftover from planet formation may also contribute to atmospheric mass loss, a mechanism known as core-powered mass loss (CPML [4]).

   Mass loss is the leading hypothesis invoked to explain the "radius valley" observed in the size distribution of exoplanets between sub-Neptunes, with radii ($R_p$) of 1.8 – 4 Earth-radii ($R_E$), and super-Earths, with $R_p$ = 1–1.8 $R_E$. The valley is thought to signal the boundary between those planets that retain hydrogen/helium atmospheres accreted from their birth environment and those that do not [5].

   It is important to differentiate whether PE or CPML is the dominant mass loss mechanism because each predicts dramatically different compositions for the residual atmospheres, even for planets that end up with identical masses and radii [6]. Using observations of the planetary transit duration in the blue wing of the Hydrogen I Ly-α emission line (1216 Å) and simultaneous monitoring of the stellar FUV radiation (1203 Å - 1560 Å), UV-SCOPE will help determine if the main cause of mass loss in planets is PE or CPML [7].

   The stellar FUV observations will be used to predict the extreme UV (EUV; 400 Å - 912 Å) radiation that drives photoevaporation but cannot be observed directly. For each star, UV-SCOPE measures the emission line fluxes of H I (Ly-α), C II, Si IV, C IV and the FUV continuum. We will use the stellar atmosphere code PHOENIX [8] to compute semi-empirical non-local thermodynamic equilibrium (non-LTE) model spectra guided by the flux in these emission lines. PHOENIX has atomic and molecular data suitable for the conditions characteristic of the stellar layers where FUV and EUV fluxes originate and has been very successful in modeling main sequence stars and predicting flux at EUV wavelengths [9].

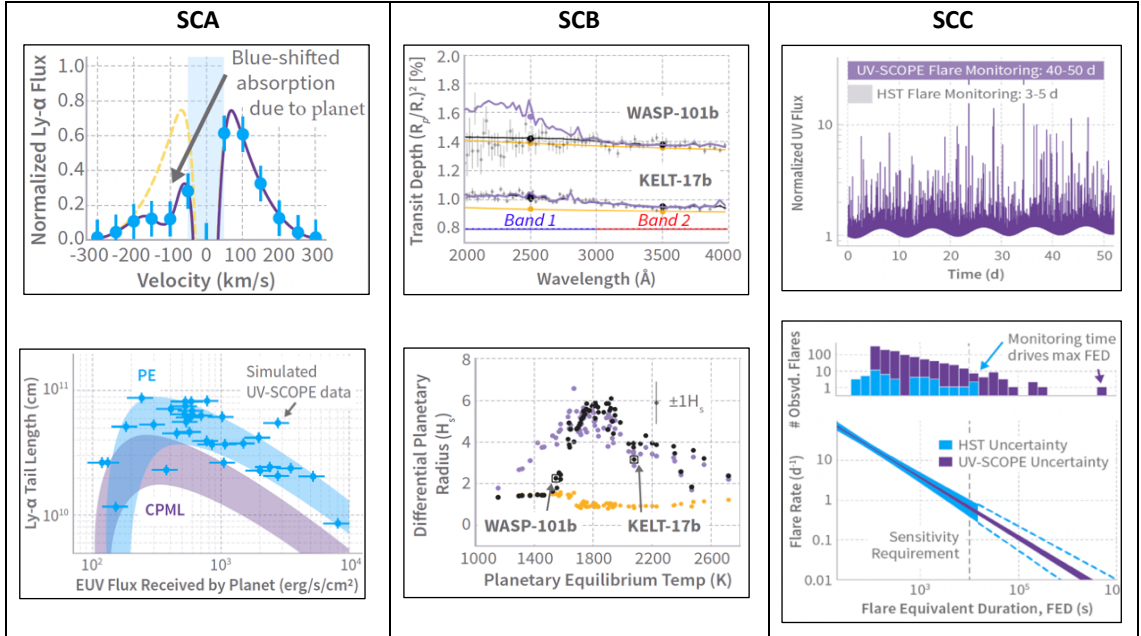

**Figure 3.** Summary of observables for each science case. *SCA - Left-Top*: When transiting the star, Hydrogen outgassing from the planetary atmosphere will be detected as absorption in the short-wavelength (blue) side of the stellar Ly-α emission line. *Left-Bottom*: When paired with information on the stellar EUV flux, the observations can discriminate between different evaporation mechanisms (PE or CPML). *SCB - Center*: Planet transmission spectroscopy in the NUV spectral range will be used to distinguish atmospheres dominated by rainout chemistry (*black dots and black lines*), pure equilibrium (*purple*) and pure Rayleigh scattering (*yellow*). The differential planetary radius between *Band 1* and *Band 2* is sensitive to these different types of chemistry. The plots highlight potential results for two representative targets. *SCC - Right-Top*: UV-SCOPE represents a large qualitative increase in the amount of UV time-domain observations of stellar variability. *Right-Bottom*: Such extreme coverage will allow UV-SCOPE to measure many more flares and reconstruct the flare frequency and energy distribution with much higher precision than has been possible thus far.

B. Science Case B (SCB): What is the composition of the upper atmospheres in hot Jupiters?

UV-SCOPE will carry-out the first systematic campaign to test theories of aerosol formation in highly irradiated Jovian exoplanets.

Aerosols (defined here as clouds and photochemical hazes) are ubiquitous in exoplanet atmospheres, but aerosol formation is poorly understood. The presence or absence of aerosols alters the temperature structure and chemistry of exoplanet atmospheres. The interpretation of most *JWST* and *ARIEL* exoplanet spectra will be uncertain because of the uncertainties in aerosol properties.

The hot Jupiters (1000 K to 2600 K) in the UV-SCOPE sample are part of a population between the cold, Solar System giants and the hot inner-heated non-irradiated brown dwarfs and young, self-luminous planets. They provide a unique laboratory for testing the hypotheses on aerosol formation; the results of the UV-SCOPE study are thus extensible to the entire planet population, including those in our Solar System.

Transit observations of these hot Jupiters, obtained in the spectral range between 2000 and 4000 Å, will sample aerosol precursors such as SiO, Mg, and Fe in order to test the leading hypotheses for the onset of aerosol formation within the hot Jupiter population: equilibrium condensation, rainout condensation, and photochemical haze production [10, 11, 12]. Equilibrium and rainout condensation arise due to the condensation of gas-phase refractory species into solid or liquid droplets as atmospheric temperatures cool, much like water clouds on Earth, but with far more exotic species. Photochemical production arises due to intense UV radiation driving the destruction of otherwise stable gaseous reservoirs into complex soots and is expected to result in a smooth Rayleigh-like continuum spectrum.

C. Science Case C (SCC): What influence does the stellar environment have on a planet's atmospheric evolution, chemistry, and habitability?

The abundance of greenhouse and biosignature molecules in planetary atmospheres is determined by photochemical, geological, and biological processes. Of these, photochemistry is the most readily accessible to astronomers through observations of the input stellar UV emission. To identify signatures of geology and biology actively modifying a planet atmosphere from below, the first step is to quantify the UV radiation modifying it from above. For terrestrial planets, UV-driven photochemistry affects the planet's climate, influencing its capacity to host life. At the same time, UV-driven photochemistry can lead to false positives and false negatives in the remote detection of life [13].

Atmospheric chemistry is highly sensitive to modest variability in UV radiation due to: UV evolution as stars spin down with age (i.e., a drop by 2–3 orders of magnitude in flux over billion-year timescales), activity cycles (factor of a few over decade timescales), rotational variability (10–30% over hour-to-month timescales) and flares (2–15,000× over second-to-hour timescales) [13, 14, 15]. Past studies of stellar UV environments have been limited to specific stellar types, with inconsistencies in wavelength coverage and minimal flare monitoring of a day or less.

Through a dedicated survey, UV-SCOPE will characterize the variability of stellar UV emission, comprehensively and consistently. These data will reveal bulk properties and dynamics of stellar upper atmospheres across age and spectral type. Plasma emission measures and densities throughout the chromosphere and transition-region will be probed by several strong FUV lines and will help identify particle acceleration fluxes. UV-SCOPE's monitoring will expand the number of flares known to exhibit hot (>$10^4$ K) FUV continua by ~ 10×, while providing simultaneous measurements of the NUV Balmer-jump crucial to probing the origin of this hot emission.

Stellar upper-atmosphere models constrained by UV-SCOPE spectra will be used to reconstruct the stellar EUV emission. For any known and yet-to-be-discovered planet, UV-SCOPE's survey will enable an informed exploration of plausible histories of that planet's NUV, FUV, and EUV irradiation, incorporating flares, rotation, activity cycles, and age-evolution.

UV-SCOPE will significantly advance the state of exoplanet research in all these science areas, leading to new discoveries and deeper questions that will guide the development of the next generation of missions.

## 3. INSTRUMENT

### 3.1 Design and Performance

During the design process, science requirements were translated into measurement requirements using design reference mission simulations of mission performance, considering the baseline target list, and the instrument radiometric model. This process resulted in a baseline instrument that fulfills the science requirements with significant margin.

The optical diagram of the payload is shown in Figure 4, and the instrument parameters are summarized in Table 1. Light from the 60-cm two-mirror telescope is focused on the center of a field stop, collimated by a third mirror, illuminates a LiF prism, and it is focused by a fourth mirror onto two delta-doped EM-CCD detectors, resulting in the spectra shown in Figure 1. All the mirrors are coated in aluminum, protected by $MgF_2$.

The field stop, or slit, blocks the light of other objects in the field, but it is large (45" spatial x 9" spectral) compared to the size of the Point Spread Function Full Width at Half Maximum (PSF FWHM=0.9"). The large size minimizes losses due to pointing errors and thermoelastic changes in the system. In the spatial direction, empty regions will be used to measure the astrophysical and instrumental background in order to subtract it from the target spectrum during processing.

Pointing requirements are driven by the need to maintain a high enough resolution to distinguish the blue lobe from the rest of the Ly-α line, for a variety of targets. Systematic errors of ≤ 200 parts-per-million in the transit depth determination are included in all calculations, but given the large slit width, this assumption does not drive pointing requirements. Visit-to-visit repeatability ensures that the target star can be placed near the same location on the spectrometer focal plane. Coarse guidance sensors in the spacecraft (S/C) point the observatory to an initial accuracy of 5″ per axis. To achieve the required stability (≤ 0.07") and repeatability (≤ 0.9"), a guidance focal plane array (FPA) of the same type as the science FPAs is placed next to the field stop. Pointing angles from this guidance FPA, derived from the position of field stars, are provided to the S/C Attitude Control System at 5 Hz, enabling fine pointing of the observatory.

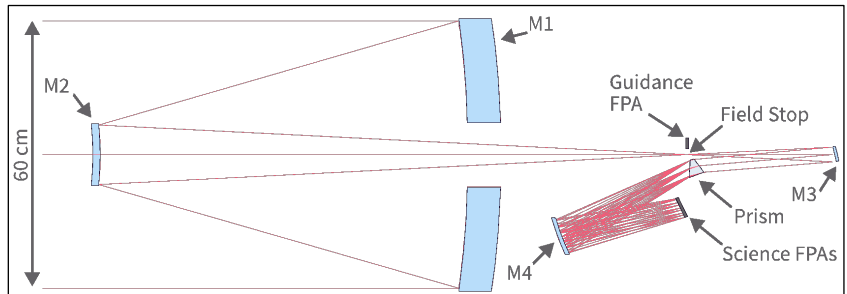

**Figure 4.** Payload optical diagram. The very simple optical path results in an efficient, high throughput system. The two-mirror telescope focuses light on the center of a field stop (slit). M3 and M4 are collimators. The guidance FPA is used to provide fine guidance pointing information to the S/C. The radiation-protection shielding surrounding the instrument is not shown.

**Table 1.**

| Parameter | Expected Value |
|---|---|
| Aperture diameter | 60 cm |
| F# | 10 |
| Slit Size | 45" x 9" |
| Spectral Range | 1203 Å - 1560 Å; 1570 Å - 4000 Å |
| Spectral resolution | 5500 at 1216 Å; 2600 at 1560 Å; 240 at 4000 Å |
| Detector type | Science: 2 x Teledyne-e2V CCD 201-20<br>Guiding: 1 x Teledyne-e2V CCD 201-20 |
| Detector size | 1056 x 2069; 13 μm pixels |

The instrument resolution is designed to satisfy the observing requirements. As we have mentioned, at 1216 Å, those requirements stem from the need to separate the blue lobe from the rest of the Ly-a line. At longer FUV wavelengths, the resolution is driven by the need to separate from each other the fluxes of lines used as input for the EUV models. In the NUV, the resolution is determined by the need to separate the molecular features that contribute to aerosol formation.

The system has been designed to minimize the number of reflective and transmissive surfaces, in order to maximize throughput over a broad spectral area (Figure 5). The resulting system has a FUV throughput (4% to 6%) comparable to the maximum throughput reached by the Cosmic Origins Spectrograph on the Hubble Space Telescope (*HST/COS*), but over a larger spectral range (1203 Å to 1560 Å vs. ~1200 Å to 1450 Å for *HST/COS*). UV-SCOPE's NUV throughput is 5 to 10 times larger than the Space Telescope Imaging Spectrograph on *HST* (*HST/STIS*) [16, Fig 5]. Compared to the Galaxy

Evolution Explorer (*GALEX*) imaging modes, UV-SCOPE's effective area is 7 times larger in the FUV and 8 times larger in the NUV.

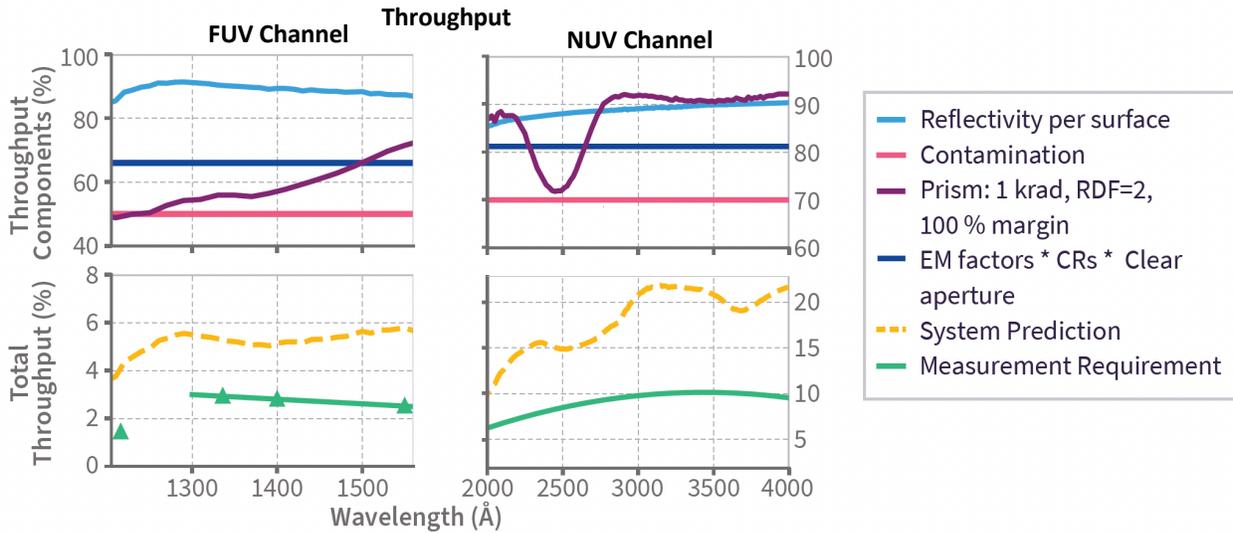

**Figure 5.** UV-SCOPE performance meets requirements with margin. The reflectivity per surface corresponds to Al-MgF$_2$ coating. Contamination control procedures are expected to maintain throughput losses at <50% (FUV) and <30% (NUV). For the prism throughput, we assume a radiation dose of 1 krad-Si, 4 times larger than expected over the 3 years in space required for the primary mission. Also shown are factors associated with the operation of the EM-CCD (Section 3.3), the impact of cosmic rays (CRs), and the effect of the secondary and its structure. Detector QE is shown in Figure 6.

### 3.2 The Prism as Dispersive Element

Using a prism instead of a grating has several key advantages for UV-SCOPE: the prism throughput is larger (Figure 5) and dichroics and order sorting filters are not needed, which avoids having additional surfaces and subsequent light losses. In the case of UV-SCOPE, the science requires high resolution at short wavelengths, and low resolution at long wavelengths and this matches well with the prism's dispersion curve.

LiF prisms and lenses are used in other space missions. Most notably, a LiF prism is the dispersive element of *HST*'s Advanced Camera for Surveys (*ACS*) Solar-Blind Channel (1150 Å to 1700 Å), which has been operating without significant degradation since ACS's installation in 2001 [17]. LiF lenses are used within NIRCam (0.5 μm to 6 μm), currently flying on *JWST* [18].

Exposing refractive elements to long-term radiation can change the real and imaginary parts of the material's index of refraction. For LiF, this is expected to be most pronounced in the NUV spectral range. To determine the suitability of LiF for UV-SCOPE, the team collaborated with the Radiation Effects Group at the Jet Propulsion Laboratory (JPL) to perform experiments in which multiple crystal samples were exposed to several increasing Co-60 radiation doses for a total exposure above and beyond that expected in the mission. Here we highlight the result of exposure to 1 krad-Si Total-Ionizing Dose (TID) of radiation, corresponding to 4x the expected value for 3 years in space. The throughput curve developed the well-known F-center at 2500 Å [e.g., 19], resulting in a 20% throughput reduction, and the expected small degradation in the FUV. These reductions are incorporated in the assumed end-of-life (EOL) performance of the instrument (Figure 5), although the actual EOL performance is likely to be better. The change in the prism dispersion is negligible.

In addition to the potential long-term changes, the interaction between the prism and solar energetic particles (SEPs) or Galactic Cosmic Rays (GCRs) results in Cherenkov radiation, fluorescence, and phosphorescence. These effects are collectively known as luminescence [20]. The contribution of luminescence to the background budget depends on the L2 environment, the instrument shielding, the straylight paths within the instrument, and the quality of the specific LiF prism to be used for flight.

Although UV-SCOPE is expected to operate during the descending slope of Solar Cycle 25, we evaluated the impact of luminescence assuming the worst possible radiation environment: solar maximum for SEPs, and solar minimum for GCRs. Combined with the system's optomechanical configuration, we concluded that the dominant luminescence contribution is fluorescence. Following [21] we measured the fluorescence conversion efficiency as part of the same radiation experiments described above. The conclusion of the analysis is that increased luminescence will result in at most 1 week per year in which the background for the faintest targets will be above the requirements. Discarding the data affected by these events results in a negligible impact on the overall efficiency of the observatory.

### 3.3 Detectors and Electronics

A particular challenge of spectroscopy is to design systems that provide a large response over a broad spectral range. UV-SCOPE uses two delta-doped EM-CCD detectors [22, 23] coated with single-layer AR coatings: $AlF_3$ and $Al_2O_3$ for the FUV and NUV, respectively. As shown in Figure 6, the resulting Quantum Efficiency (QE) is generally larger, over a broad spectral range, than the values reached by Micro-Channel Plate (MCP) devices [24].

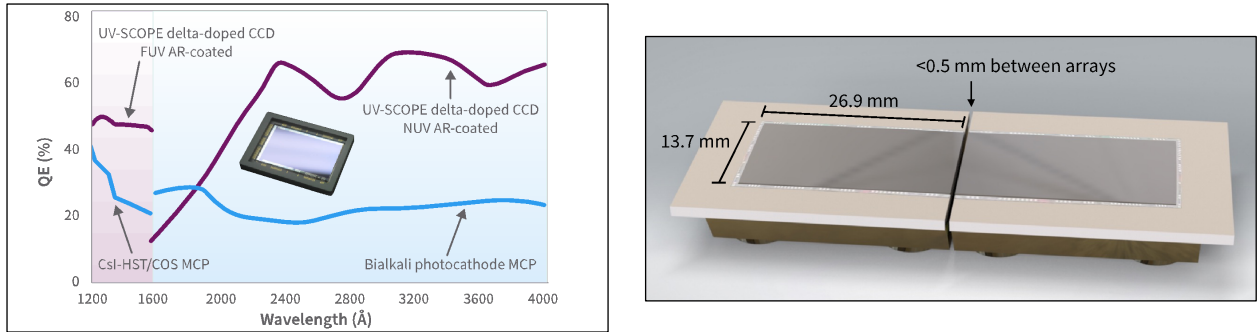

**Figure 6.** UV-SCOPE uses two 1K x 2K delta-doped EM-CCD detectors, coated with single-layer AR coatings, and with a small separation between them, to image spectra between 1203 and 4000 Å, with a 10 Å gap beyond 1560 Å. The panel on the left compares the detector QE with state-of-the-art MCPs [24]. Stray light due to target radiation emitted at wavelengths longer than 4000 Å is controlled by a strip of black silicon [25], beyond the sensing area of the detectors (not shown).

The spectra will be laid down along the long axis of the two detectors. Because the prism dispersion decreases with wavelength, light redder than 4000 Å is confined to a short strip just beyond the sensing area of the detectors and will be absorbed by a black silicon surface in order to control stray-light [25].

The system's sensitivity is driven by the need to detect the FUV continuum of the faintest target, expected to be $7.5 \times 10^{-18}$ ergs/s/cm$^2$/Å. To achieve this limit, the FUV detector is operated in electron-multiplying, "photon counting" mode. In this mode, the signal is amplified by factors of several thousand, and the read-noise is negligible. In our calculation of the throughput, we include coincidence (5%) and detection losses (5%), associated with the operation of the EM-CCDs [23], in addition to charge transfer efficiency losses (10%).

The NUV detector is also an EM-CCD, although it will generally be operated in conventional mode, as the targets are brighter in this spectral range.

EM-CCDs are extensively used on ground facilities (e.g. the `Alopeke and Zorro instruments in Gemini Observatories [26]), have been used in flight projects (e.g. Fireball [27]), and are baselined to be used in the Roman Space Telescope [28].

The electronic system is designed to allow the detectors to be operated simultaneously but independently, with different exposure times, gains, or operational modes.

## 3.4 Spacecraft and Thermal

The S/C bus (Figure 7) will be fabricated by Ball Aerospace. It is part of Ball Configurable Platform (BCP), a single-string, small bus product line, that includes the US Department of Defense *Space Test Program* satellites (STPSat-2/-3), NASA's *Green Propellant Infusion Mission* (GPIM), and the *Wide-field Infrared Survey Explorer* (WISE, now NEOWISE) missions, as well as S/C in development for NASA's *Imaging X-ray Polarimetry Explorer* (IXPE), *Spectro-Photometer for the History of the Universe, Epoch of Reionization, and Ices Explorer* (SPHEREx), *Global Lyman-alpha Imagers of the Dynamic Exosphere* (GLIDE), and *Space Weather Follow On-Lagrange 1* (SWFO-L1) missions. The S/C TID tolerance at L2 is >25 krad, exceeding the expected contribution of SEPs and GCRs. The only consumable is the hydrazine propellant for the thrusters in charge of station-keeping. The S/C will carry enough propellant for over 6 years of operations.

Thermal control is primarily passive, supplemented with operational and survival heaters, all using standard Ball design techniques. The instrument optics and spectrometer are isolated from the bus via the hexapod struts using low conductance brackets and bolts at the mounting interface on the S/C deck.

A passive cryogenic radiator, mounted on the anti-sun side of the S/C and operating at 163 K, is used to cool the science arrays to < 168 K, limiting dark current to a negligible level. Heat is conducted from the science detectors to the radiator by pyrolytic graphite flexible thermal links. The mechanical supports of the science arrays and cryogenic radiator are thermally isolating. The optics operate at 220±10 K, passively cooled through the telescope aperture (also at 220 K). The outer barrel assembly (OBA) is wrapped in multi-layer insulation (MLI) to thermally isolate it from the surroundings and limit spatial gradients. The instrument electronics are mounted inside the OBA with a dedicated radiator (at 300 K) that can tolerate expected solar flux.

If the concept is approved, the system will be built over a period of 5 years. Travel to L2 will take approximately 2 months, and the primary science mission will take 34 months to be executed.

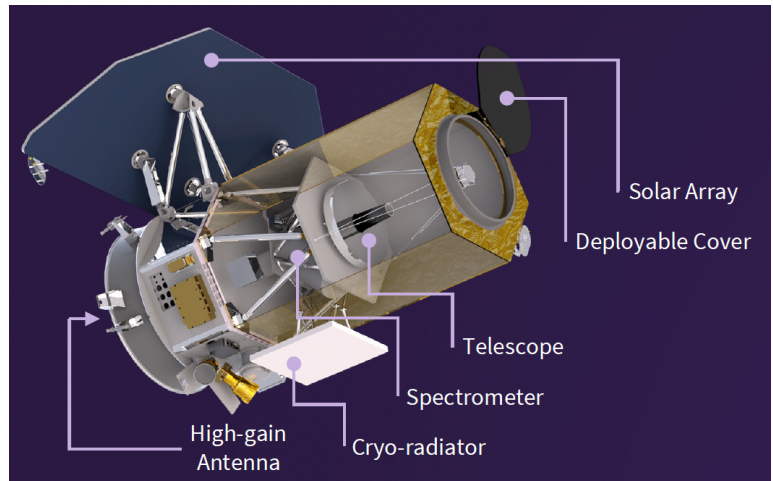

**Figure 7.** UV-SCOPE uses Ball's BCP S/C, which has ample heritage in space. The telescope is surrounded by a hexagonal barrel assembly, to protect the UV telescope during launch and limit stray-light contamination. The payload is passively cooled via a 0.4 m$^2$ radiator. 4 monopropellant thrusters are used to provide station keeping in L2, and their fuel is the only consumable in the system. The deployable cover is the only moving part in the mechanical system. It is expected to be opened soon after launch and remain open for the duration of the mission.

## 4. CONCLUSION

UV-SCOPE is a mission concept to study the radiation environments of exoplanets and their atmospheres. It was proposed to NASA as MidEx mission. If approved, the system will be built over 5 years, and then launched to L2 for a 34-month science mission.

The instrument design is simple and therefore robust. UV-SCOPE will have the only instrument, past, present, or planned, that provides near-continuous single-shot 1203–4000 Å spectra, with throughputs comparable to *HST/COS* and 10× better than *HST/STIS*.

Realistically, UV-SCOPE science questions cannot be answered with current facilities. For example, the survey contemplated in SCB would take over 9 years if performed with *HST* at the exclusion of all other *HST* science. The survey described in the context of SCC would take over 7 years with *HST*.

While its primary *Worlds and Suns in Context* spectroscopic survey has enduring legacy value, UV-SCOPE will also have value as a general observing facility, after this primary mission. UV-SCOPE's spectroscopic capability will have applications in many areas, including Solar System planet aurorae, star formation, protoplanetary disks, planet magnetic fields, massive-star winds, active galactic nuclei, and gravity-wave follow-up. In addition, UV-SCOPE provides opportunities for time-domain programs with a dedicated observatory that has no Earth occultations or South Atlantic Anomaly interference.

UV-SCOPE continues the on-going trend in exoplanet science, transitioning from the era of "planet hunting" to an era of "planet characterization," focusing on planet atmospheres. This is a crucial step towards answering the most fundamental question of all: "Are we alone?"

## 5. ACKNOWLEDGMENTS

UV-SCOPE is a collaboration between Arizona State University, Jet Propulsion Laboratory / California Institute of Technology, Ball Aerospace, the Laboratory for Atmospheric and Space Physics, the Infrared Processing and Analysis Center, Eureka Scientific, NASA's Goddard Space Flight Center, Imperial College London, The Johns Hopkins University Applied Physics Laboratory, University of Arizona at Tucson, University of California at Berkeley, University of California at Los Angeles, University of California at Santa Cruz, Utah Valley University, and the University of Washington at Seattle.

The UV-SCOPE team recognizes the support of the JPL Astronomy and Astrophysics Program Office, in the funding of the concept study.

The research was carried out at the Jet Propulsion Laboratory, California Institute of Technology, under a contract with the National Aeronautics and Space Administration (80NM0018D0004). ©2022. All rights reserved.